\documentclass[pra,twocolumn,eqsecnum,showpacs,amsmath,amssymb,superscriptaddress]{revtex4-1}
\usepackage{graphicx,bm,color,mathptmx,hyperref} 
\usepackage[latin9]{inputenc}
\usepackage{color}
\usepackage{amsmath}
\usepackage{amssymb}
\usepackage{graphicx}
\usepackage{esint}
\usepackage{braket}

\makeatletter
\begin{document}

\title{Efficient single photon absorption by a trapped moving atom}

\author{N. Trautmann}

\affiliation{Institut f\"ur Angewandte Physik, Technische Universit\"at Darmstadt,D-64289,
Germany}

\author{G. Alber}
\affiliation{Institut f\"ur Angewandte Physik, Technische Universit\"at Darmstadt,D-64289,
Germany}

\author{G. Leuchs}

\affiliation{Max-Planck-Institute for the Science of Light, G\"unther-Scharowsky-Stra{\ss}e
1, Bldg 24, D-91058 Erlangen, Germany}

\affiliation{Institute of Optics, Information and Photonics, Department of Physics,
University Erlangen-N\"urnberg, Staudtstra{\ss}e 7/B2, D-91058 Erlangen,
Germany}


\date{\today}
\begin{abstract}
The influence of the center of mass motion of a trapped two level system
on efficient resonant single photon absorption is investigated.  It is shown that this absorption process depends strongly on the ratio between the
characteristic time scales of spontaneous photon emission and of the two level system's center of mass motion. In particular, if the spontaneous photon emission process occurs almost instantaneously on the time scale of the center of mass motion coherent control of the center of mass motion offers interesting perspectives for optimizing  single photon absorption.
It is demonstrated that this way time dependent modulation of a harmonic trapping 
frequency
allows to squeeze the two level system's center of mass motion so strongly that high efficient single photon absorption is possible even in cases of weak confinement by a trapping potential\footnote{This work was initiated by Roy J. Glauber who wanted to see a proper
theoretical treatment of an experimentally viable scenario for
demonstrating the process, which is the time reversed version of the
spontaneous emission process of an atom. We acknowledge his
advice and guidance.}.
\end{abstract}


\maketitle

\section{Introduction}

Recent technological advances in the area of resonant photon-matter interaction have opened new exciting experimental possibilities \cite{Berman1994,Walther2006,Haroche2006}. 
One line of research in this direction focuses on the development of
efficient means for coupling a single elementary quantum system, such as a trapped atom or ion, to properly engineered multimode radiation fields in order to achieve a controlled and almost perfect
transfer of excitation between few photon multimode states and material quantum systems \cite{fischer2013efficient,lindlein2007new}. Besides being of fundamental quantum optical interest
these investigations are also driven by the desire to explore new possibilities for realizing efficient ways of transferring quantum information between material elementary two level systems (stationary qubits) and photons (flying qubits) \cite{PhysRevLett.114.173601}.

Recently it has been demonstrated experimentally that by trapping a single elementary quantum system in the center of a parabolic mirror an efficient coupling to optical photonic multimode states can be achieved \cite{fischer2013efficient,lindlein2007new,Maiwald2012}. Thereby, a parabolic mirror constitutes a convenient tool for redirecting and focusing an
asymptotically incoming (almost) plane wave 
containing a few photons onto an elementary material quantum system trapped in the parabola's focus.  For focal lengths and distances large in comparison with the wave lengths of optical photons
the quantized radiation field dominantly coupling to such a trapped material quantum system looks like a dipole field in free space whose center is located in the focus of the parabola. Guided by characteristic features of
the spontaneous photon emission process it has been demonstrated theoretically that even in free space optimal single photon wave packet states exist which are capable of exciting an elementary two level system at a fixed position in space almost perfectly.
However, the center of mass motion of an absorbing material quantum system complicates the situation considerably because in general single photon absorption and the subsequent spontaneous photon emission process together with the resulting recoil effects \cite{PRA.55.R2539} entangle the center of mass and the photonic degrees of freedom in an intricate way.
On the basis of these general features it appears that achieving almost perfect single photon excitation in free space or with the help of a parabolic mirror in the presence of center of mass motion requires preparation of a highly entangled quantum state of the center of mass and the photonic degrees of freedom. Even by nowadays technological capabilities the controlled preparation of such entangled
quantum states constitutes a major technological obstacle and appears unrealistic.

Motivated by these developments we explore the influence of the center of mass motion of a trapped material two level system on single photon absorption in free space  or equivalently in a parabolic cavity with large focal length. In particular, we explore possibilities for optimizing this process with the help of the particular single photon wave packet \cite{Stobinska2009} which would achieve almost perfect absorption in the absence of any center of mass motion.
At first sight one may be tempted to conclude that almost perfect
excitation of a trapped two level system by such an optimal single photon wave packet is only achievable in sufficiently strongly confining traps with large trap frequencies. However, our investigations demonstrate that almost perfect
photon absorption from such an optimal single photon wave packet is possible even  in weakly confining traps and for initially prepared thermal states of the center of mass motion provided the trap frequency is modulated periodically in an appropriate way. Our analysis exhibits also the crucial role played by the characteristic dynamical parameters, namely the spontaneous photon emission rate of the electronic transition involved and the trap frequency.
The investigation of the impact of the center of mass motion on atom field interactions for high-finesse cavities has recently also attracted attention in the literatur, see \cite{PhysRevA.85.013830} for example.

This paper is organized as follows. In Sec. \ref{quantum electrodynamical model} the quantum electrodynamical model describing single photon absorption by a trapped moving two level system is presented. Based on the dipole and rotating wave approximation the dynamics of the single photon excitation process and its relation to the relevant field correlation function is discussed in section \ref{Dynamics}. In Sec. \ref{SmallLD} characteristic features of the excitation probability and its deviation from the ideal motionless case are investigated for tightly confining trapping potentials.
In the subsequent section \ref{ArbitraryLD} these results are generalized to weakly confining trapping potentials with particular emphasis on the experimentally interesting dynamical regime of spontaneous photon emission rates large in comparison with relevant trapping frequencies. Sec. \ref{CoherentCM} finally explores possibilities to optimize the excitation probability by periodic modulation of the trap frequency of a harmonic trap. 

\section{A quantum electrodynamical model\label{quantum electrodynamical model}}

\begin{figure}
\begin{center} 
\includegraphics[width=0.45\textwidth]{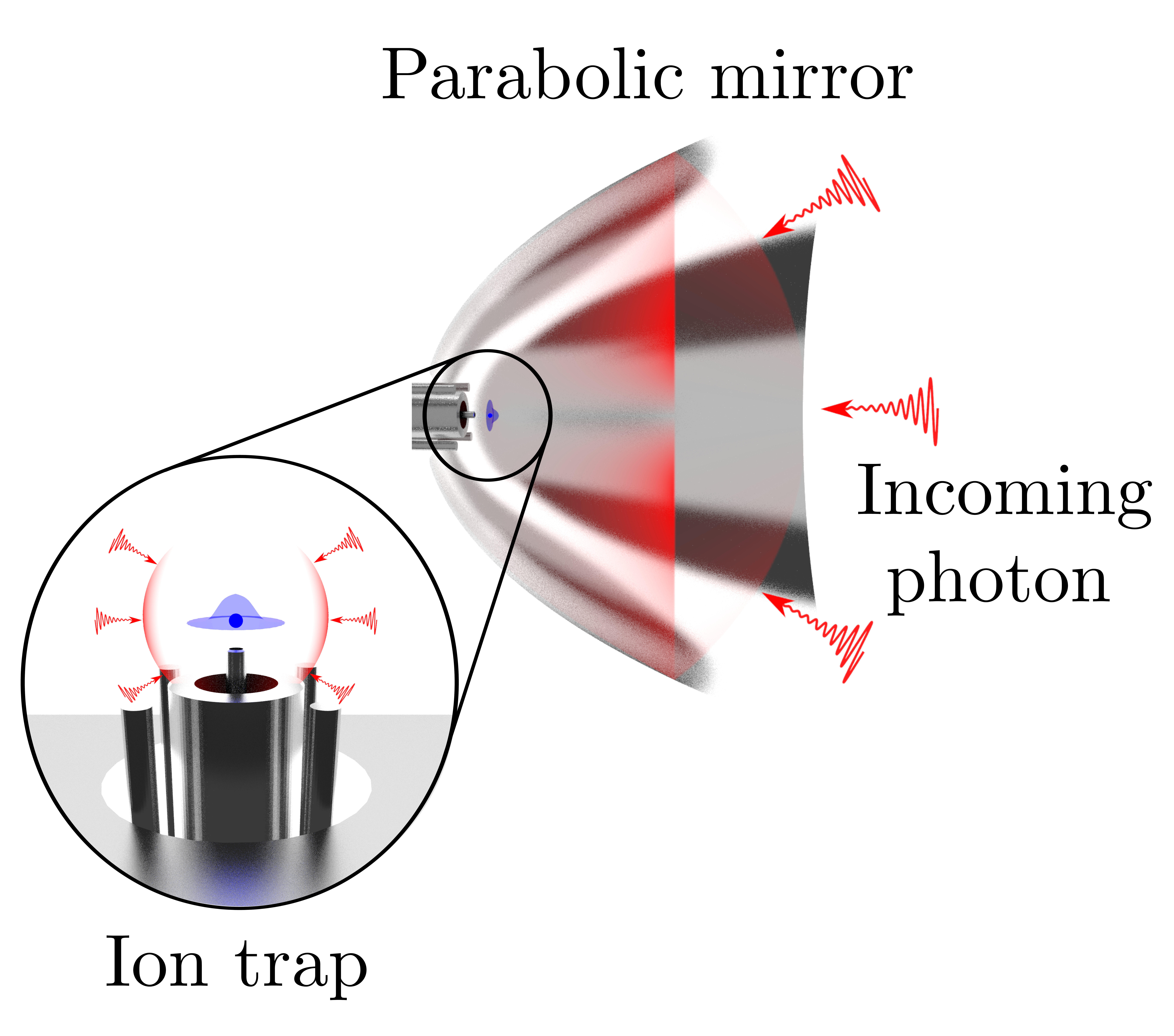}
\end{center} 
\caption{
Single photon absorption process in a parabolic cavity: A two level
system trapped close to the focal point ${\bf x}=0$ of a parabolic
cavity is excited by an asymptotically incoming single photon wave
packet. This wave packet couples to the two level system in the same
way as a single photon wave packet capable of exciting this two level
system almost perfectly in free space (see Eq.(\ref{inputfield})). A suitable setup with a trapped
ion at the focal point of the parabolic mirror has been described
in \cite{fischer2013efficient,Maiwald2012,Maiwald2009}
\label{fig:Setup}}
\end{figure}

We consider a single trapped atom or ion, whose internal electronic dynamics is modeled by a two level system resonantly coupled to an optical  single photon radiation field and explore it's capability of absorbing this single photon almost perfectly.
Inspired by recent experiments \cite{fischer2013efficient,lindlein2007new,Maiwald2012} we assume that the center of the trap is positioned in the focal point of a large parabolic mirror which is capable of focusing a well directed asymptotically incoming single photon radiation field towards the two level system trapped close to the focal point of the parabola.
This excitation process is depicted in Fig. \ref{fig:Setup} schematically.

In order to describe the dynamics of this single photon excitation process we take advantage of the fact that in the  optical frequency regime  typical wave lengths
are large in comparison with the Bohr radius characterizing spatial extensions of atoms or ions in energetically low lying bound electronic states. Therefore, the interaction between the two level system and the radiation field can be described in the dipole approximation.
Furthermore, we assume that the time scale of typical optical transitions is orders of magnitude smaller than all other interaction-induced time scales, i.e. the optical transitions frequency $\omega_{eg}$ is large in comparison to the spontaneous decay rate in free space
$\Gamma$. For atoms or ions interacing with the radition field in the optical frequency regime this is well satisfied, see for example the parameters of the experiment described in \cite{Maiwald2012} (with $\Gamma=1.2\cdot 10^8 \text{s}^{-1}$ and $\omega_{eg}=5.1\cdot  10^{15} \text{s}^{-1}$ ).
In this regime the rotating wave approximation is applicable.
In addition, the velocity of the two level system's center of mass motion is assumed to be negligible in comparison with the speed of light in vacuum  $c$
so that the dynamics of the center of mass motion can be described in the non relativistic approximation. Under these conditions 
the total Hamiltonian governing the dynamics of the quantum electrodynamical interaction between the moving two level system and the radiation field is given by
%
\begin{equation}
\hat{H}=
\hat{H}_{\text{E}}+\hat{H}_{\text{T}}+\hat{H}_{\text{F}}+\hat{H}_{\text{int}}.\label{eq:Hamiltonian}
\end{equation}
The free evolution of the radiation field is described by the Hamiltonian
$\hat{H}_{\text{F}}=\hbar\sum_{i}\omega_{i}\hat{a}_{i}^{\dagger}\hat{a}_{i}$
with the photonic destruction and creation operators $\hat{a}_i$ and $\hat{a}_i{\dagger}$ and with $i$ indexing the modes with frequency $\omega_i$ of the transverse radiation field.
The Hamiltonians $\hat{H}_{\text{E}}= E_g |g\rangle \langle g| + E_e |e\rangle \langle e|$ and $\hat{H}_{\text{T}}={\hat{\mathbf{p}}^{2}}/{2m}+{V}_{\text{T}}(\hat{\mathbf{x}})$ describe the dynamics of the electronic degrees of freedom and of the
center of mass of the trapped two level system.  
The frequency of the relevant electronic transition
is denoted $\omega_{eg} = (E_e - E_g)/\hbar$
and
$\hat{\mathbf{x}}$, $\hat{\mathbf{p}}$, and $m$ are position operator,
momentum operator, and mass of the two level system's center of mass degrees of freedom. The Hamiltonian
$\hat{H}_{\text{int}}=-\hat{\mathbf{E}}^{(+)} (\hat{\mathbf{x}})\cdot{\mathbf{d}}|e\rangle \langle g| +\text{H.c.}$ describes the interaction between the two level system and the transverse radiation field in the dipole and rotating wave approximations with the two level system's dipole matrix element ${\mathbf{d}} = \langle e|\hat{\mathbf{d}}|g\rangle$ and its dipole operator
$\hat{\mathbf{d}}$. The mode decomposition of the positive part of the transverse electric field operator is given by
\begin{eqnarray}
\hat{\mathbf{E}}^{(+)} ({\mathbf{x}}) &=&
\sum_j i\sqrt{\frac{\hbar \omega_j}{2\epsilon_0}} {\bf g}_j({\mathbf{x}}) \hat{a}_j.
\end{eqnarray}
The orthonormal transverse mode functions
${\bf g}_i({\mathbf{x}})$ are solutions of the Helmholtz equation with appropriate boundary conditions and with unit normalization per field mode.
($\epsilon_0$ is the dielectric constant of the vacuum.)
%

In order to describe the dynamics of the absorption by a single photon we have to solve the time dependent Schr\"odinger equation with Hamiltonian
(\ref{eq:Hamiltonian}). In particular, we are interested in its solution with a separable pure state of  the form
\begin{equation}
\ket{\psi_{{0}}}=\ket{g}\ket{\psi_{\text{T}}}\ket{\psi_{\text{F}}}
\label{initial}
\end{equation}
prepared initially at time $t=t_{0}$.
Thereby, $\ket{\psi_{\text{T}}}$ describes the pure initial state
of the center of mass degree of freedom and $\ket{\psi_{\text{F}}}$ denotes
the initial one photon state of the radiation field. 
In the following we are particularly interested in the center of mass induced dynamics of this single photon excitation process for a photon state which is capable of exciting a two level system located at the fixed position ${\bf x} =0$, such as the focal point of a parabolic mirror, almost perfectly.

\section{Dynamics of single photon absorption\label{Dynamics}}

In this section we explore the dynamics of optimal resonant single photon absorption (in free space) by a moving trapped two level system whose Hamiltonian is given by Eq.(\ref{eq:Hamiltonian}).

In view of the rotating wave approximation the Hamiltonian (\ref{eq:Hamiltonian})
conserves the numbers of excitations. Therefore, for an initial state of the form of Eq. (\ref{initial}) the general solution of the Schr\"odinger equation is given by a pure quantum state which is a superposition of the photonic vacuum state correlated with the excited two level system and of
a single photon multimode state correlated with the two level system in its ground state and with its center of mass degrees of freedom generally entangled with the field modes.
As the center of mass motion is assumed to be non relativistic the resulting modification of the spontaneous photon
emission process from the excited two level system is negligible so that it is still described by the spontaneous decay rate $\Gamma$ in free space, i.e.
\begin{eqnarray}
\Gamma &=& \frac{\omega_{eg}^3\mid\mathbf{d} \mid^2}{3\pi \epsilon_0 \hbar c^3}\;.
\end{eqnarray}
Thus, from the time dependent Schr\"odinger equation we find that
 the probability of detecting at time $t$ the two level system in its excited state $|e\rangle$ is given by
\begin{eqnarray}
P_e(t) &=&
\int_{t_0}^{t} dt_1
e^{i(\omega_{eg} +i \Gamma/2)(t-t_1) }
\int_{t_0}^{t} dt_2
e^{-i(\omega_{eg} -i \Gamma/2)(t-t_2) }~~~~\label{P}\\
&&
\frac{{\mathbf{d}}^*}{\hbar}\cdot
\langle \psi_{T}| 
\mathbf{G}^{(1)}(\hat{\mathbf{x}}_I(t_1),t_1),\hat{\mathbf{x}}_I(t_2),t_2))
|\psi_{T}\rangle
\cdot\frac{{\mathbf{d}}}{\hbar}.\nonumber
\end{eqnarray}
This excitation probability is determined by the mean value of the normally ordered field correlation tensor of first order, i.e.
\begin{small}
\begin{eqnarray}
\mathbf{G}^{(1)}(\hat{\mathbf{x}}_I(t_1),t_1,\hat{\mathbf{x}}_I(t_2),t_2)
&=&
\langle \psi_{F}|
\hat{\mathbf{E}}^{-} (\hat{\mathbf{x}}_I(t_1),t_1)\otimes
\hat{\mathbf{E}}^{+} (\hat{\mathbf{x}}_I(t_2),t_2)
|\psi_{F}\rangle,\nonumber
\end{eqnarray}
\end{small}
averaged over the two level system's initially prepared center of mass state $|\psi_T\rangle$. Thereby, the position
operator 
\begin{eqnarray}
\hat{\mathbf{x}}_I(t) &=& \exp(i\hat{H}_{T}(t-t_0)/\hbar)~
\hat{\mathbf{x}} ~
\exp(-i\hat{H}_{T}(t-t_0)/\hbar)\nonumber
\end{eqnarray}
denotes the time evolution of the center of mass position of the two level system in the interaction picture. In this interaction picture the transverse electric field operators $\hat{\mathbf{E}}^{(+)} ({\mathbf{x}}_I(t),t)$ and
$\hat{\mathbf{E}}^{(-)} ({\mathbf{x}}_I(t),t) = \left(\hat{\mathbf{E}}^{(+)} ({\mathbf{x}}_I(t),t)\right)^{\dagger}$ are defined by
\begin{eqnarray}
\hat{\mathbf{E}}^{(+)} ({\mathbf{x}},t) &=&
\sum_j i\sqrt{\frac{\hbar \omega_j}{2\epsilon_0}} {\bf g}_j({\mathbf{x}}) e^{-i\omega_j(t-t_0)}\hat{a}_j.
\end{eqnarray}

Let us now concentrate on an initially prepared one photon state $|\psi_F\rangle$ which achieves almost perfect excitation of a two level system positioned at the fixed position ${\bf x} = 0$  in free space. It has been demonstrated \cite{Stobinska2009} that the pure single photon state
\begin{eqnarray}
|\psi_F\rangle &=& \sum_j \hat{a}_j^{\dagger}|0\rangle
\frac{1}{\hbar}\sqrt{\frac{\hbar \omega_j}{2\epsilon_0}}
\frac{
{\mathbf{d}}^* \cdot {\bf g}_j({\bf x}=0)}{\omega_j-\omega_{eg} - i \Gamma/2}
e^{i\omega_j \left(t_{\text{out}}-t_{0}\right)}~~~~
\label{inputfield}
\end{eqnarray}
prepared at time $t_0$ 
with $\Gamma (t_{out} - t_0) \gg 1$
achieves such an almost perfect excitation at time $t_{\text{out}}$
in the absence of any center of mass motion. 
The frequencies $\omega_i$ of this one photon state are distributed according to a Lorentzian
spectrum centered resonantly around the two level system's transitions frequency $\omega_{eg}$. The relative phases between these modes are determined by the parameter $t_{out}$ which describes
the time at which a two level system at the fixed position ${\bf x} = 0$ is excited
almost perfectly.
If the two level system were positioned in the center of a spherically symmetric cavity
of radius $R$ with ideally conducting walls, for example, the discrete orthonormal 
mode functions coupling to the two level system in the dipole approximation would be given by
{\begin{eqnarray}
{\bf g}_n({\bf x})&=& 
\sqrt{\frac{1}{R}} \nabla\wedge \left(
j_1(\omega_nr/c)~ {\bf x}\wedge
\nabla Y_1^0(\theta,\varphi) \right)= 
\label{modefunction}
\\
&&
-\sqrt{\frac{3}{4\pi R}}\frac{\omega_n}{c}\left(
e^{i\omega_n r/c} {\bf g}_{\omega_{n}}^{(+)}({\bf x}) +
e^{-i\omega_n r/c} {\bf g}_{\omega_{n}}^{(-)}({\bf x}) 
\right)\nonumber
\end{eqnarray}
with
\begin{eqnarray}
{\bf g}_{\omega}^{(+)}({\bf x}) &=&
-{\bf e}_r \cos \theta \left(
\frac{1}{(\omega r/c)^2}
+\frac{i}{(\omega r/c)^3}
\right) -\\
&&
{\bf e}_{\theta}
\frac{\sin \theta}{2}
\left(
\frac{1}{(\omega r/c)^2}
-i\left(\frac{1}{(\omega r/c)} -
\frac{1}{(\omega r/c)^3} 
\right) 
\right)\nonumber
\end{eqnarray}
and
with the angular momentum eigenfunction
$Y_1^0(\theta,\varphi)$ \cite{AS}.
For real-valued frequencies $\omega$ the relation
${\bf g}_{\omega}^{(-)}({\bf x}) = \left({\bf g}_{\omega}^{(+)}({\bf x})\right)^*$ applies.
Here, $j_1(u) = \sin u/u^2 -\cos u/u$ denotes the spherical Bessel function of fractional order \cite{AS} and
$r=\mid {\bf x}\mid$. The angle between the direction of the dipole matrix element 
$\mathbf{d}$
and ${\bf x}$ is denoted $\theta$. Furthermore, ${\bf e}_r$ and ${\bf e}_{\theta}$ are  the spherical coordinate
unit vectors in the $r$ and $\theta$ directions
with the $z$-direction oriented parallel to the dipole vector $\mathbf{d}$.
In the continuum limit of  large cavity sizes, i.e. $\omega_{eg} R/c \gg 1$,
the mode frequencies are given approximately by
$\omega_n = c\pi(n+1)/R$ with $n\geq 0$ being integer. 

In accordance with current experimental activities
\cite{lindlein2007new,fischer2013efficient,Maiwald2012}
and with the scenario depicted in  Fig. \ref{fig:Setup} this excitation process can be realized with the help of a parabolic mirror with large focal length $f$ and with the two level system's trap centered in the parabola's focal point ${\bf x} =0$. 
In the geometric optical limit in which $f$ is large in comparison with the photon's  wave lengths the optimal single photon state of Eq.(\ref{inputfield}) can
be prepared by focusing an appropriately polarized (almost) plane wave
single photon state which asymptotically enters the parabola along the direction of its symmetry axis
\cite{Alber2013}.

For the pure one photon state $|\psi_F\rangle$ the first order correlation tensor factorizes, i.e. 
\begin{eqnarray}
\mathbf{G}^{(1)}(\hat{\mathbf{x}}_I(t_1),t_1,\hat{\mathbf{x}}_I(t_2),t_2)
&=& \left({\mathbf{F}^{*}}(\hat{\mathbf{x}}_I(t_1),t_1)\right)\otimes
{\mathbf{F}}(\hat{\mathbf{x}}_I(t_2),t_2)\nonumber
\end{eqnarray}
with the effective one photon operator $
{\mathbf{F}}(\hat{\mathbf{x}}_I(t),t) = \langle 0|
\hat{\mathbf{E}}^{(+)}(\hat{\mathbf{x}}_I(t),t)|\psi_F\rangle$.
For our initial state, we get
\begin{eqnarray}
 \mathbf{F}({\bf x},t)&=&\frac{i}{2\epsilon_{0}}\sum_{n}\omega_{n}{\bf g}_{n}(\mathbf{x})e^{-i\omega_{n}(t-t_{0})}\frac{\mathbf{d}^{*}\cdot{\bf g}_{n}({\bf x}=0)}{\omega_{n}-\omega_{eg}-i\Gamma/2}e^{i\omega_{n}\left(t_{\text{out}}-t_{0}\right)}\nonumber\\
&=&\frac{i}{\epsilon_{0}}\sum_{n}\frac{\omega_{n}^{3}}{4\pi Rc^{2}}\left(e^{-i\omega_{n}t_{+}}{\bf g}_{\omega_{n}}^{(+)}({\bf x})+e^{-i\omega_{n}t_{-}}{\bf g}_{\omega_{n}}^{(-)}({\bf x})\right)\nonumber\\
&&\cdot\frac{\mathbf{d}^{*}\cdot e_{z}}{\omega_{n}-\omega_{eg}-i\Gamma/2} \end{eqnarray}
with $t_{\pm} = t - t_{\text{out}} \mp r/c$.
In the continuum limit, i.e. $R\to~\infty$, we can replace the sum by an integral.
We obtain
\begin{eqnarray}
\mathbf{F}({\bf x},t)&\underset{R\rightarrow\infty}{=}&\frac{i}{4c^3\epsilon_{0}\pi^2}\int_{0}^{\infty}d\omega\left(e^{-i\omega t_{+}}{\bf g}_{\omega}^{(+)}({\bf x})+e^{-i\omega t_{-}}{\bf g}_{\omega}^{(-)}({\bf x})\right)\nonumber\\
&&\cdot\frac{\omega^{3}\mathbf{d}^{*}\cdot e_{z}}{\omega-\omega_{eg}-i\Gamma/2}\;.
\end{eqnarray}
By extending the region of integration to $-\infty$ (which is well justified for $\omega_{eg}\gg\Gamma$) and by applying Cauchy's residue theorem,
we obtain the following analytical expression
\begin{eqnarray}
{\bf F}({\bf x},t) &=&
-\frac{\hbar\Gamma3}{{\bf d}\cdot {\bf e}_z2}\left[
{\bf g}_\omega^{(+)}({\bf x}) \Theta(-t_+) e^{-i\omega t_+}+\right.\nonumber\\
&&\left.
{\bf g}_\omega^{(-)}({\bf x}) \Theta(-t_-) e^{-i\omega t_-}
\right]\mid_{\omega = \omega_{eg} + i\Gamma/2}
\label{One-photon-general}\;.
\end{eqnarray}
Close to the ideal position
${\bf x} = 0$ of the two level system, i.e. for $\eta = \omega_{eg} r/c\ll 1$,
this one photon operator simplifies to 
\begin{eqnarray}
{\mathbf{F}}(\mathbf{x},t)&=& - \frac{\hbar \Gamma}{\mathbf{d}\cdot {\bf e}_z}
e^{i(\omega_{eg} + i \Gamma/2)(t_{\text{out}} - t)}\Theta(t_{\text{out}} - t)\times~~\label{F}\\
&&
\left(
{\bf e}_z
\left(1 - \frac{\eta^2}{10}\right)+
{\bf e}_{\theta}\sin\theta \frac{\eta^2}{10}  
+ O(\eta^ 4)
\right).\nonumber
\end{eqnarray}
Inserting Eq.(\ref{One-photon-general}) into Eq.(\ref{P})
yields the time dependence of
the excitation probability $P_e(t)$.
For a two level system fixed at position ${\bf x} = 0$
and for times
$t\geq t_0$
this one photon excitation probability reduces to
\begin{eqnarray}
P _{id}(t)&=& e^{-\Gamma\mid t_{\text{out}} - t\mid }\left( 1 - e^{-\Gamma(\tau - t_0)}\right)^2
\end{eqnarray}
with $\tau = t \Theta(t_{\text{out}} - t) +  t_{\text{out}} \Theta(t - t_{\text{out}})$.
Thus, for large interaction times, i.e. $\Gamma(\tau - t_0)\gg 1$,
the single photon state of Eq.(\ref{inputfield})
achieves almost perfect excitation of the two level system at time
$t_{\text{out}}$ apart from terms exponentially small in the parameter
$\Gamma(\tau - t_0)\gg 1$.

Depending on the ratio between the characteristic time scale
of the center of mass motion and of the single photon absorption and spontaneous emission process
two extreme dynamical cases can be distinguished.
If the trap frequency $\omega_T$ of a harmonically trapped two level system
is much larger than the spontaneous decay rate $\Gamma$
the spontaneous decay process is so slow that details of the center of mass motion are
averaged out in the time integrals of Eq.(\ref{P}). Consequently, the excitation probability dominantly
depends on  the time averaged center of mass motion. In the opposite limit,
i.e. $\omega_T \ll \Gamma$, the spontaneous photon emission process occurs almost
instantaneously on the time scale of the center of mass motion. Consequently, the time
evolution of the excitation probability depends on details of the two level system's center
of mass motion in the trap.

\section{Single photon absorption and strong confinement of the center of mass motion\label{SmallLD}}
Inserting Eq.(\ref{F}) into Eq.(\ref{P})
a systematic understanding of the influence of the two level system's
center of mass motion on the single photon absorption process can be obtained in cases in which this motion
is confined to a region close to the ideal position ${\bf x} = 0$ in the sense that
$\eta = \omega_{eg} r/c\ll 1$. 
From Eqs.(\ref{F}) and (\ref{P})we obtain the result
\begin{small}
\begin{eqnarray}
&&P_e(t) =\Gamma^2
\int_{t_0}^{t} dt_1~ 
\int_{t_0}^{t} dt_2 ~
\Theta(t_{\text{out}} - t_1)
\Theta(t_{\text{out}} - t_2)
e^{-\Gamma (t+t_{\text{out}}-t_1 -t_2)}\times\nonumber\\
&&
\left( 
1 -
2\frac{\omega^2_{eg}}{10c^2}
\langle \psi_{T}|
\left(
\hat{z}^2_I(t_2) + 2\hat{x}^2_I(t_2)+2\hat{y}^2_I(t_2)\right)|\psi_{T}\rangle
+ O(\eta^4)
\right)
\label{PLD}
\end{eqnarray}
\end{small}
with $\hat{x}_I(t), \hat{y}_I(t),\hat{z}_I(t)$
denoting the time dependent Cartesian $x,y,z$ components of the position operator
of the center of mass degrees of freedom in the interaction picture. 
Thereby, the $z$-direction is oriented parallel to the dipole vector $\mathbf{d}$. 

Let us investigate the center of mass motion in an anisotropic harmonic trapping potential
of the form 
\begin{eqnarray}
{V}_T ({\bf x})&=&m\omega_z^2 z^2/2 + m \omega_x^2 x^2/2 + m \omega_y^2 y^2/2 
\label{potential}
\end{eqnarray}
in more detail.
In the
interaction picture the resulting dynamics of the $z$
component of the center of mass position
is given by
\begin{eqnarray}
\hat{z}_I(t) &=& \hat{z} \cos(\omega_z (t-t_0)) + 
\frac{\hat{p_z}}{m\omega_z}\sin(\omega_z (t-t_0))  
\label{z}
\end{eqnarray}
with analogous expressions for the other Cartesian components.
Position and momentum operators in the Schr\"odinger picture are
denoted by $\hat{z}$ and $\hat{p}_z$ etc.. 
Inserting these position operators into Eq.(\ref{PLD})
yields the result
\begin{eqnarray}
\frac{P_e(t)}{P_{id}(t)} &=&
1 - \frac{\omega_{eg}^2}{5c^2}
\left(
\frac{\hbar A_z(\tau)}{2m\omega_z} +
2\sum_{j=x,y}\frac{\hbar A_j(\tau)}{2m\omega_j}
\right)
\label{strongconf}
\end{eqnarray}
with  
$\tau = t \Theta(t_{\text{out}} - t) +  t_{\text{out}} \Theta(t - t_{\text{out}})$
and with
\begin{eqnarray}
A_j(\tau) &=&
\langle \psi_T | \hat{b_j}^{\dagger 2}
| \psi_T\rangle
\frac{\Gamma e^{2i\omega_j (\tau - t_0)}}{\Gamma + 2i\omega_j}
\frac{1 - e^{-(\Gamma + 2i\omega_j)(\tau - t_0))}}{1 - e^{-\Gamma(\tau - t_0)}} + 
\nonumber\\
&&
\langle \psi_T | (\hat{b_j}^{\dagger}\hat{b_j} + 1/2)
| \psi_T\rangle + c.c.
\end{eqnarray}
for $j=x,y,z$.
The creation and destruction operators of the harmonic oscillators in the Cartesian directions $j=x,y,z$ are denoted by
$\hat{b}_j^{\dagger}$ and
$\hat{b}_j$. 

According to Eq.(\ref{strongconf}) for long interaction times, i.e. $\tau - t_0 \gg 1/\Gamma$, and for small spontaneous photon emission rates,
i.e. $\Gamma \ll \omega_j$, the excitation probability $P_e(t)$ is determined by  the time averaged center of mass
motion. In this dynamical regime we have $A_j(\tau) = 2 \langle \psi_T | (\hat{b_j}^{\dagger}\hat{b_j} + 1/2)|\psi_T\rangle + O(\Gamma/\omega_j)$ so that the deviation of the excitation probability $P_e(t)$ from its ideal value $P_{id}(t)$  is proportional to the mean energy of the center of mass degrees of freedom in the harmonic trap, i.e.
\begin{eqnarray}
\frac{P_e(t)}{P_{id}(t)}
 &=& 
1 - \frac{\omega_{eg}^2}{5c^2}
\left(
\frac{\hbar \omega_z  \langle \psi_T | (\hat{b_z}^{\dagger}\hat{b_z} + 1/2)|\psi_T\rangle}{m\omega^2_z} +\right.\nonumber\\
&&
\left.
2\sum_{j=x,y}\frac{\hbar \omega_j \langle \psi_T | (\hat{b_j}^{\dagger}\hat{b_j} + 1/2)|\psi_T\rangle}{m\omega^2_j}
\right).
\end{eqnarray}
As the mean energy of a harmonic oscillator is lower bounded by its zero point energy in this dynamical regime the excitation probability 
$P_e(t_{out})$ assumes its maximal value if the center of mass degrees of freedom are prepared
in the ground state of the harmonic trap so that $\hat{b}_j|\psi_T\rangle = 0$ for $j=x,y,z$.

According to Eq.(\ref{strongconf}) in the opposite limit of long interaction times, i.e. $\tau - t_0 \gg 1/\Gamma$, but large spontaneous photon emission rates,
i.e. $ \Gamma \gg \omega_j$, the excitation probability $P_e(t)$ is determined by the center of mass
motion at time $\tau = t \Theta(t_{\text{out}} - t) + t_{\text{out}}\Theta(t - t_{\text{out}}) $. In this dynamical regime we obtain the approximate result
\begin{eqnarray}
A_j(\tau)  &=&
\langle \psi_T |
\left(
 \hat{b_j}^{\dagger} e^{i\omega_j (\tau - t_0)}
 + 
\hat{b_j} e^{-i\omega_j (\tau - t_0)} 
\right)^2
| \psi_T\rangle
\end{eqnarray}
by neglecting terms of the order of $O(\omega_j/\Gamma)$.
The resulting strong dependence of the excitation probability $P_e(t)$ on the center of mass motion  at time $\tau$ can be exploited for
minimizing the deviations of the excitation probability from its ideal values $P_{id}(t)$. This can be achieved by preparing appropriate initial states $|\psi_T\rangle$ which minimize the quantities $A_j(\tau)$.
For this purpose
squeezed vacuum states \cite{squeezedstates} of
the center of mass motion of the form
\begin{eqnarray}
|\psi_T\rangle &=&
\hat{S}(\xi_z) \hat{S}(\xi_x)\hat{S}(\xi_y)|0_T\rangle
\label{initialstate}
\end{eqnarray}
 offer interesting possibilities.  In Eq.(\ref{initialstate})
 $|0_T\rangle$ denotes the ground state of the harmonic trap, i.e. $\hat{b}_j|0_T\rangle = 0$ for $j=x,y,z$ and
\begin{eqnarray}
\hat{S}(\xi_j) &=& e^{\xi_j^* \hat{b}_j^2 /2 - \xi_j \hat{b}_j^{\dagger 2} /2}
\end{eqnarray}
are the squeezing operators for the
Cartesian components $j$.
For complex valued squeezing parameters
of the form $\xi_j = r_j {\rm exp}(2i\varphi)$ with $r_j> 0$,
for example, we obtain
the result
\begin{eqnarray}
\frac{P_e(t)}{P_{id}(t)} &=&
1 - 
\frac{\omega_{eg}^2}{5c^2} \left(
\Delta z_0^2 e^{-2r_z}
+ 2 \Delta x_0^2 e^{-2r_x}
+ 2 \Delta y_0^2 e^{-2r_y}
\right).\nonumber\\
\label{Psqueezed}
\end{eqnarray}
at all times $t$ for which
$\sin(\omega_j(\tau - t_0)-\varphi) = 0$. Thereby, the quantities $\Delta z_0 =  \sqrt{\hbar/(2 m \omega_z)}$ etc. denote
the extensions of the ground state of the harmonic trap along 
the corresponding Cartesian directions.
In contrast,
at times for which $\cos(\omega_j(\tau - t_0)-\varphi) = 0$
these extensions are enhanced periodically by
corresponding factors of
$e^{2r_z}$ etc.
Thus,
strong squeezing of
the initial state of Eq.(\ref{initialstate}) in all
directions, i.e. $r_j\gg 1$,
implies a significant increase of the excitation probability
at 
all interaction times $t$ with
$\sin(\omega_j(\tau - t_0)) = 0$
and
may lead to significantly more efficient single photon excitation
than achievable in cases of small spontaneous decay rates. 

\section{Single photon absorption and weak confinement of the center of mass motion\label{ArbitraryLD}}

If the center of mass motion is not confined to a region
around the ideal position
${\bf x} =0$ small in comparison with the wave length of
the resonantly absorbed photon a Taylor expansion
of
the mode functions ${\bf g}_n({\bf x})$
around
${\bf x} = 0$ is no longer adequate. In this case we start
from the general form of the 
one photon operator
as given by
Eq. (\ref{One-photon-general}).
Inserting this 
expression
into Eq. (\ref{P}) yields the excitation probability.

For typical spontaneous decay times $1/\Gamma$ of the order of nanoseconds
and extensions of center of mass wave packets
$\Delta x$ small in comparison with a few meters so that
$\Gamma \Delta x/c \ll 1$  this one photon operator 
can be further simplified to the expression
\begin{eqnarray}
{\bf F}({\bf x},t) &=&
-\frac{3\hbar\Gamma}{{2\bf d}\cdot {\bf e}_z}\Theta( t_{\text{out}} -t) e^{-i(\omega_{eg} + i\Gamma/2) (t-t_{\text{out}})}\nonumber\\
&&\left( {\bf g}_n^{(+)}({\bf x}) e^{i\omega_n r/c}+\right.
\left.
{\bf g}_n^{(-)}({\bf x})e^{-i\omega_n r/c} 
\right)\mid_{\omega_n = \omega_{eg}}.\nonumber\\
\end{eqnarray}

In many trapping experiments
typical spontaneous photon emission rates
of electronic transitions $\Gamma$ are large in comparison with
typical trap frequencies, i.e. $\Gamma \gg \omega_T$.
In this dynamical regime the single photon absorption process takes
place almost instantaneously on the characteristic time scale induced by the trapping potential.
In particular, we demand that for the typical velocities $\Delta v(t)=\sqrt{\langle \psi_T|\left(\frac{d}{dt}\mathbf{\hat{x}}_{I}(t)\right)^2|\psi_T\rangle}$ the condition
\begin{eqnarray}\Delta v(t)/\left(\lambda_{eg}\Gamma\right)\ll1\label{Doppler_condition}\end{eqnarray}
is fulfilled within a time interval of the order of a few lifetimes of the excited state,
i.e. $1/\Gamma$, around $t_{\text{out}}$ ($\lambda_{eg}$ is the wave length of the optical
transition $|e\rangle \to |g\rangle$.). For a thermal state of the harmonic trap, we obtain
\begin{equation}
\Delta v(t)=\sqrt{\langle\hat{H}_{j}\rangle/m}\;\text{for }j\in\left\{ x,y,z\right\}
\end{equation}
with $\hat{H}_{j}$ denoting the Hamiltonian describing the dynamics of the center of mass motion along the $j$-axis.
By using this result , we get for the ground state
\begin{equation}\Delta v(t)=\sqrt{\omega_{j}\hbar/\left(2m\right)}\;\text{for }j\in\left\{ x,y,z\right\}.\end{equation}
By using the experimental parameters given in \cite{Maiwald2012} (trapping of 
$^{174}\text{Yb}$, $\omega_T=2\pi\cdot 480\text{kHz}$, $\Gamma=1.2\cdot 10^8 \text{s}^{-1}$, and $\lambda_{eq}=369\;\text{nm}$)
we obtain
$\Delta v(t)/\left(\lambda_{eg}\Gamma\right)=5.3\cdot 10^{-4}\ll1$ for the ground state
and $\Delta v(t)/\left(\lambda_{eg}\Gamma\right)=3.3\cdot 10^{-3}\ll1$ for the thermal state in the Doppler limit.
Hence, the condition is well satisfied for typical experimental parameters.
Condition (\ref{Doppler_condition}),  can also be seen as a condition
for a negligible Doppler shift in comparison with the spontaneous decay rate of the atomic transition.
Thus, the integration over times $t_1$ and $t_2$ appearing in Eq.(\ref{P})
can be performed with the help of partial integration \cite{Bender-Orszag1999} and
the excitation probability at time $t$ is determined dominantly by the
position operators
$\hat{\mathbf{x}}_I(t_1)$ and $\hat{\mathbf{x}}_I(t_2)$ 
at time
$\tau = t \Theta(t_{\text{out}}-t) + t_{\text{out}}\Theta(t - t_{\text{out}})$ 
and at the initial time $t_0$.
Consequently, for large interaction times, i.e. $\Gamma(\tau - t_0)\gg 1$, and neglecting
terms exponentially small in this parameter
Eq. (\ref{P}) simplifies to
\begin{eqnarray}
\frac{P_{e}(t)}{P_{id}(t)}&=&
\frac{9}{4}
\langle\psi_{T}|
\left(
{\bf e}_{z}\cdot{\bf g}_{n}^{(+)}(\mathbf{\hat{x}}_{I}(\tau))
e^{i\omega_{n}\mid \mathbf{\hat{x}}_{I}(\tau)\mid/c} + 
\right. \nonumber\\ &&\left.
{\bf e}_{z}\cdot{\bf g}_{n}^{(-)}(\mathbf{\hat{x}}_{I}(\tau))
e^{-i\omega_{n}\mid\mathbf{\hat{x}}_{I}(\tau)\mid/c}
\right)^{2}
\mid_{\omega_{n}=\omega_{eg}}|\psi_{T}\rangle.\nonumber\\
\label{eq:P_t}
\end{eqnarray} 

For a spherically symmetric trapping potential
a simple analytical relation can be obtained
from Eq.(\ref{eq:P_t})
if at the observation time $t$ the center of mass degrees of freedom can be described
by an isotropic Gaussian state $\hat{\rho}_T\left(\tau\right)$ centered around ${\bf x} = 0$ 
with spatial variance $\text{Tr}\left[ \hat{\rho}_T\left(\tau\right) {\hat{\bf x}}^{2}\right]  \equiv
\text{Tr}\left[ \hat{\rho}_T\ {\hat{\bf x}}^{2}_I(\tau)\right] $, 
i.e.
\begin{equation}
\frac{P_{e}(t)}{P_{id}(t)}=
\frac{3e^{-2\eta_0^{2}}}{10\eta_0^{6}}\left(-2\eta_0^{4}-\eta_0^{2}+
\left(2\eta_0^{4}-\eta_0^{2}+1\right)e^{2\eta_0^{2}}-1\right)\label{eq:p_max}
\end{equation}
with the effective (time dependent) Lamb-Dicke parameter
\begin{eqnarray}
&\eta_0=\frac{\omega_{eg}}{c_{0}}\;\sqrt{\text{Tr}\left[ \hat{\rho}_T\ {\hat{\bf x}}^{2}_I(\tau)\right]/3}.
\label{eq:def_eta_0_gen}
\end{eqnarray}
This result can be applied to a large class of center of mass states including squeezed vacuum and thermal states (with respect to the isotropic trapping potential).
The dependence of the excitation probability $P_{e}(t)/P_{id}(t)$
on the effective Lamb-Dicke parameter $\eta_0$ is depicted
in Fig. \ref{fig:max_Prob_exciting_atom} for
$\Gamma\gg\omega_{\text{T}}$. The corresponding time dependence of the excitation probability $P_{e}(t)$ is illustrated in Fig. \ref{fig:time_Prob_exciting_atom} for several values of $\eta_0$.
It is apparent that for small effective Lamb-Dicke parameters $\eta_0$ almost perfect excitation
is achievable at time $t_{\text{out}}$.
For all effective Lamb-Dicke parameters almost
instantaneous excitation with a large spontaneous
decay rate, i.e. $\Gamma \gg \omega_T$, is more effective than excitation with a small spontaneous decay rate, i.e. $\Gamma\ll\omega_{T}$.


\begin{figure}
\begin{center}
\includegraphics[width=0.45\textwidth]{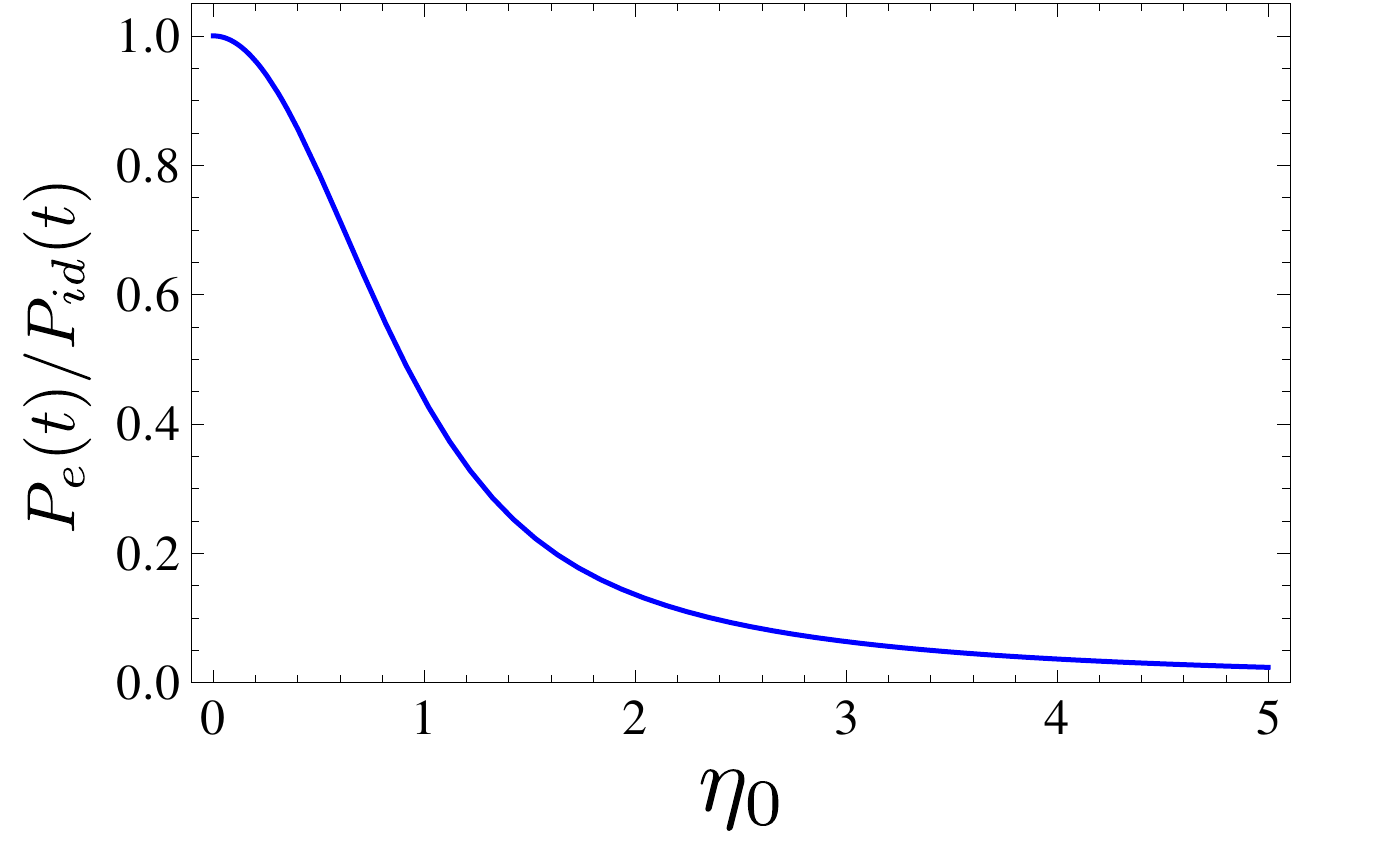}
\end{center} 
\caption{Probability 
$P_{e}(t)$ of exciting a two level system in a spherically symmetric trap
and its dependence on the effective
Lamb-Dicke parameter $\eta_0$ (see Eq.(\ref{eq:def_eta_0_gen})): The  center of mass state $\hat{\rho}_T\left(\tau\right)$ is
assumed to be an isotropic Gaussian state. The spontaneous decay rate is assumed to be large compared to the trap frequency, i.e.
$\Gamma\gg \omega_{\text{trap}}$.
\label{fig:max_Prob_exciting_atom}}
\end{figure}

\begin{figure}
\begin{center}
\includegraphics[width=0.45\textwidth]{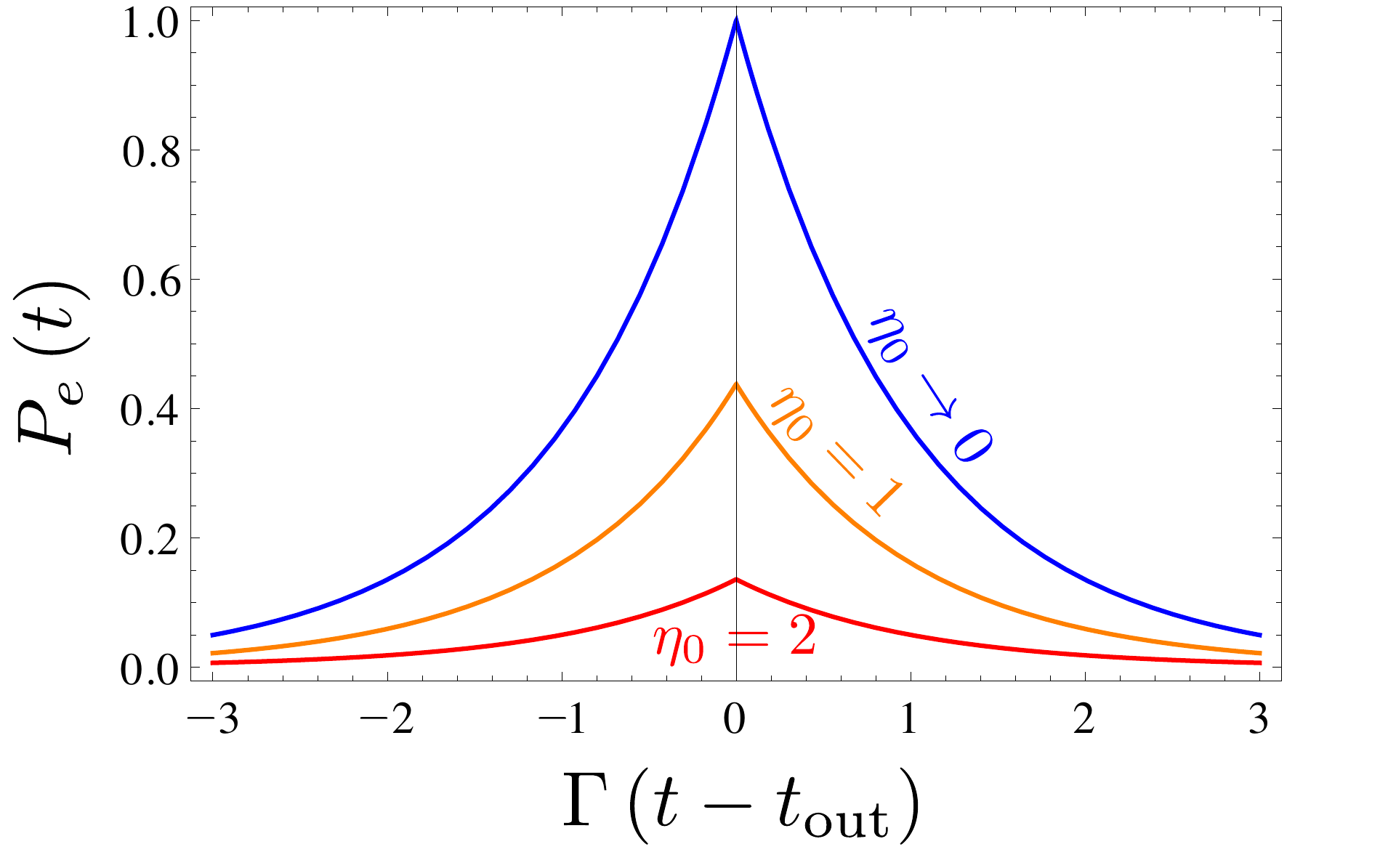}
\end{center} 
\caption{
Time dependence of
the excitation probability
$P_{e}(t)$
in a spherical symmetric trap
for an isotropic Gaussian center of mass state $\hat{\rho}_T\left(\tau\right)$:
The parameter $\eta_0$ characterizing the center of mass state is chosen to be $\eta_0\rightarrow 0$ (blue), $\eta_0=1$ (orange) and  $\eta_0=2$ (red).
The trapping frequency is small, i.e.
$\Gamma\gg \omega_{T}$, so that the time dependence of $\eta_{0}$ can be neglected. The interaction time is large, i.e.
$\Gamma (\tau -t_0)\gg1$ .
\label{fig:time_Prob_exciting_atom}}
\end{figure}

\section{Optimizing single photon absorption by coherent control
of the center of mass motion \label{CoherentCM}}
As discussed in Sec. \ref{ArbitraryLD}, the maximum probability for exciting the two level system in the regime $\Gamma\gg\omega_{\text{trap}}$ is mainly limited
by the spatial width of the center of mass state during the short time of the order $1/\Gamma$ in which the absorption process takes place. This limitation is of particular significance
in cases in which no sub-Doppler cooling techniques are applied and in which this spatial width is not sufficiently small.
A straightforward strategy to overcome this hurdle is to increase the depth of
the trap and thus increase the confinement of the center of mass degrees of freedom. This procedure, however,  is typically limited by experimental constraints  and therefore cannot constitute an ultimate solution. As discussed in Sec. IV even if the relevant trapping  frequency becomes larger than the spontaneous photon emission rate the achievable spatial confinement of the center of mass state is  limited ultimately by the zero point fluctuations in the trap.
However, as discussed in Sec. V, if  the photon absorption process takes place
almost instantaneously
it is possible to increase the excitation probability
by preparing a center of mass state
whose width is sufficiently small during  the photon absorption process, such as a squeezed state, for example.
The  squeezing of the  center of mass state in ion traps has already been demonstrated in experiment \cite{meekhof1996generation,leibfried2003quantum}.
One method for achieving significant squeezing 
is to modulate the trapping frequency with twice the trapping frequency \cite{heinzen1990quantum,schrade1997endoscopy}. 
This way highly efficient one photon excitation can be achieved even in a weakly confining harmonic trapping potential.

For this purpose let us consider the dynamics of the center of mass motion of a non
relativistic particle of mass $m$ in a periodically modulated
spherically symmetric harmonic trapping potential 
of the form
\begin{eqnarray}
V_T({\bf x},t) &=& \frac{1}{2}  m\omega^2(t) {\bf x}^2,\nonumber\\ 
\omega^2(t) &=& \omega^2_T + \omega^2_T \delta 
\sin(\omega_M(t - t_0)).
\end{eqnarray}
The trapping potential above corresponds to the well studied problem of a parametric oscillator.
The time evolution of the classical as well as the quantum mechanical problem can be expressed by using solutions of the Mathieu differential equation \cite{AS}.
It is well known that the solutions of the Mathieu equation become unstable in the region $\omega_M\approx2\omega_T$ \cite{AS}.
This phenomenon of parametric resonance can be used to achieve significant squeezing of the center of mass state.

If the modulation strength of the trapping potentials is small, i.e. $\mid\delta \mid \ll 1$, and $\omega_M=2\omega_T$  the dynamics of the center of mass motion can be determined perturbatively.
The unperturbed dynamics is defined by the modulation strength $\delta =0$ and by  the corresponding explicitly time independent Hamiltonian $\hat{H}_0 = \hat{{\bf p}}^2/(2m) + m\omega_T^2 {\bf x}^2/2$.
The resulting time evolution of an initially prepared pure state $|\psi_T\rangle$  is given by
\begin{eqnarray}
|\psi_T(t)\rangle &=&
\hat{S}(\xi_x(t))\hat{S}(\xi_y(t))\hat{S}(\xi_z(t)) e^{-i\hat{H}_0(t-t_0)/\hbar}|\psi_T\rangle
\end{eqnarray}
with the time dependent squeezing parameters

$\xi_x(t) = \xi_y (t) = \xi_z (t) = r (t) e^{-2i\varphi (t)}$
being approximately determined by
\begin{eqnarray}
 r(t) &=& \omega_T  \delta (t - t_0)/4,
\varphi (t) = \omega_T (t-t_0) - \pi/2\;.
\label{eq:xiapprox}
\end{eqnarray}

The 
time evolution of the squeezing parameter $r(t)$
is depicted in the numerical results of Fig. \ref{fig:squeezing} for several scenarios.
The plot illustrates that for $\omega_{M}=2\omega_{T}$
the numerical results are in excellent agreement with the approximate analytical expression
of Eq. (\ref{eq:xiapprox}) even for moderately large modulation amplitudes. Even if the condition
$\omega_{M}=2\omega_{T}$ is violated significant squeezing can be achieved.
For larger deviations of $\omega_M$ form  $2\omega_T$ (roughly $\mid\omega_M^2-(2\omega_T)^2\mid\gtrsim 2\delta\omega_T^2$) a transition from an unstable solution
of the Mathieu equation to a stable solution takes place.
Squeezing can also be achieved in the stable region, but in this case the value of the squeezing parameter is bounded from above (see solid blue line in Fig. \ref{fig:squeezing}).  

For an initially prepared energy eigenstate of the unperturbed trapping Hamiltonian $\hat{H}_0$
the  corresponding mean values and variances of the position operator are given by
\begin{eqnarray}
\langle \hat{\bf{x}}_I(t)\rangle &=& 0,~~
\langle \hat{\bf{x}}^2_I(t)\rangle = \frac{\langle \psi_T|\hat{H}_0 |\psi_T\rangle}{m\omega_T^2}\times\nonumber\\
&&\left(e^{-2r(t)} \cos^2 \varphi(t)  +e^{2r(t)}\sin^2\varphi(t)\right).
\end{eqnarray}
This implies that also for any incoherent mixture of energy eigenstates, such 
as a thermal state, for $\delta>0$ at times $t$ with $\sin(\varphi(t)) = 0$
the position uncertainties are squeezed significantly. 
For a thermal state the mean energy of the unperturbed isotropic
harmonic motion in the trap
at temperature $T$ is given by
$\langle \hat{H}_0 \rangle = 3\hbar \omega_T\left(1/2+1 
/[{\rm exp}(\hbar \omega_T/(kT))-1]\right)$. 
The squeezing $r(t)$ induced by the periodic modulation of the
trapping frequency with twice the trapping frequency $\omega_T$
increases linearly with the interaction time $(t - t_0)$. Thus,
it is capable of reducing
the uncertainty around the mean position
$\langle \hat{\bf{x}}_I(t)\rangle = 0$
significantly even for an initially prepared thermal state. Consequently, 
even if the center of mass motion is
confined only weakly by a trapping potential
the excitation
probability can achieve values very close to the ideal motionless case.

\begin{figure}[t]
\begin{center}
\includegraphics[width=0.45\textwidth]{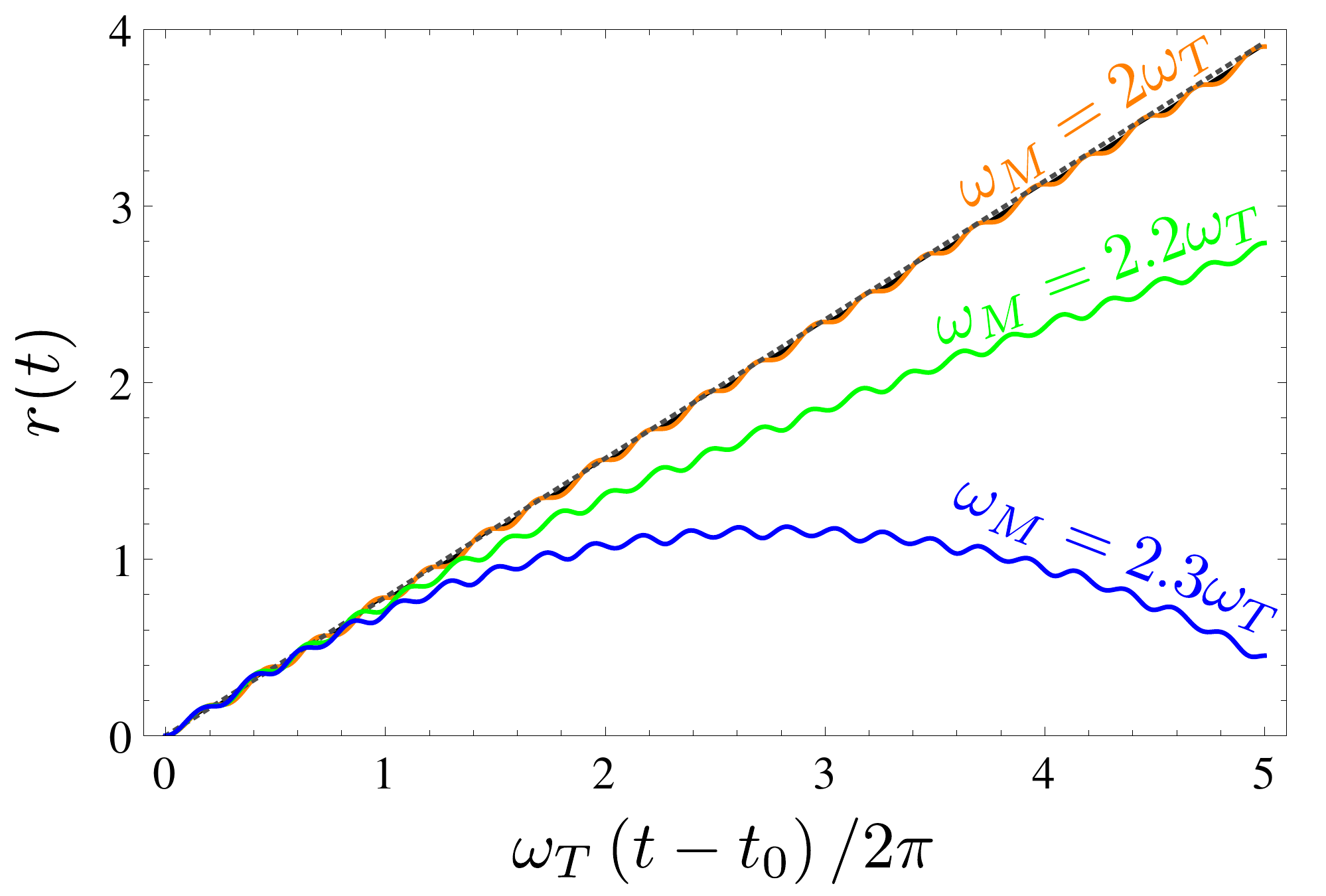}
\end{center} 
\caption{Time evolution of the squeezing parameter $r(t)$
for $\delta=0.5$:
Numerically exact results for modulation frequencies $\omega_{M} = 2 \omega_T$ (orange solid line), $\omega_{M} = 2.2 \omega_T$ (green solid line), $\omega_{M} = 2.3 \omega_T$ (blue solid line) and the
associated approximation (dotted line).
\label{fig:squeezing}}
\end{figure}

\begin{figure}[b]
\begin{center}
\includegraphics[width=0.45\textwidth]{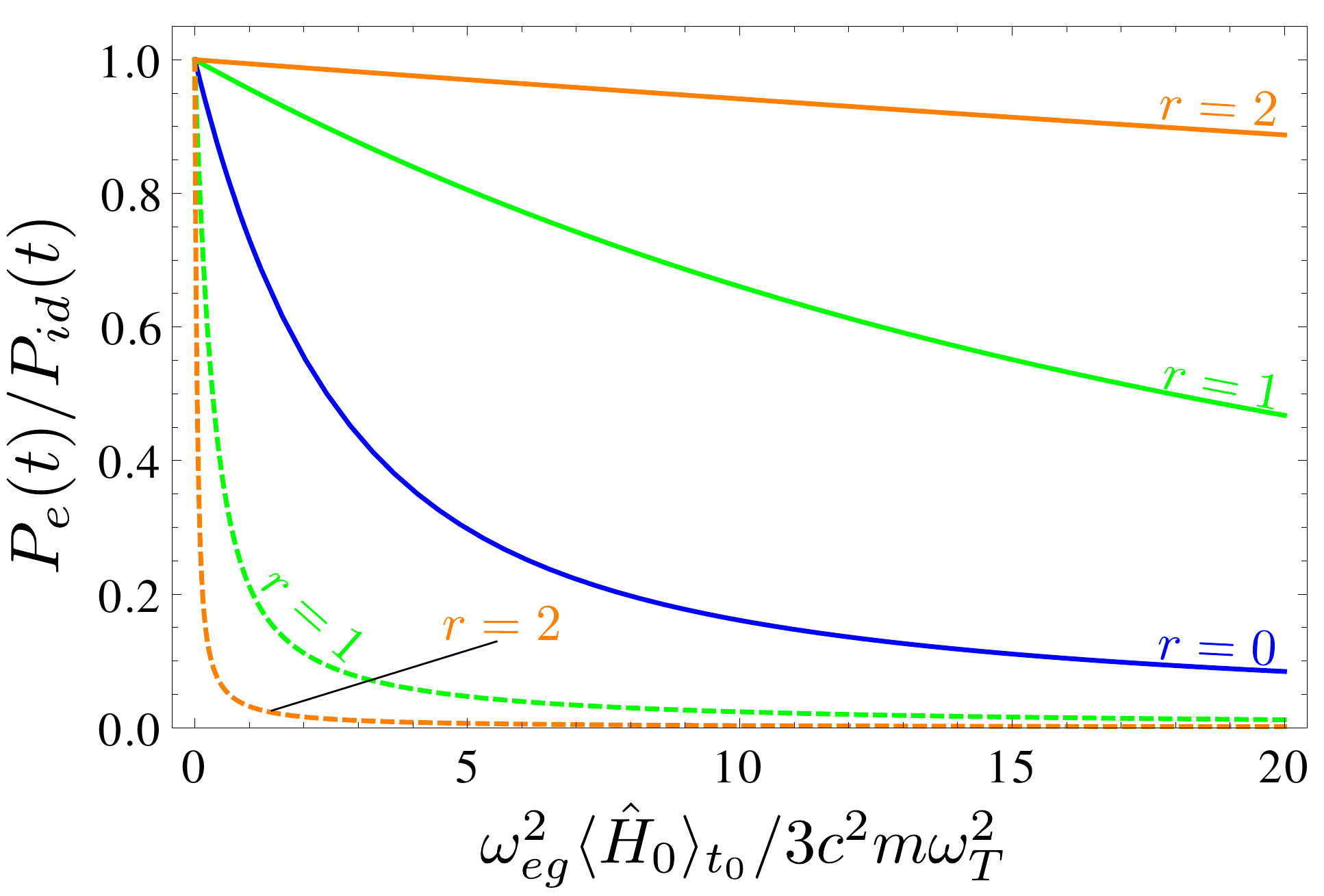}
\end{center} 
\caption{
Influence of squeezing of initially prepared thermal center of mass states on
excitation probability $P_e(t)$:
The parameters are $r=0$ (blue), $r=1$ (green) and $r=2$ (orange). The solid lines correspond to $\sin^2(\varphi(t))=0$ and the dashed lines correspond to $\sin^2(\varphi(t))=1$ during the short time the absorption of the photon takes place.
For the experimental parameters given in \cite{Maiwald2012} (trapping of 
$^{174}\text{Yb}$, $\omega_T=2\pi\cdot 480\text{kHz}$, $\Gamma=1.2\cdot 10^8 \text{s}^{-1}$, and $\omega_{eg}=5.1\cdot 10^{15}\;\text{s}^{-1}$)
we obtain $\omega_{eg}^2\langle \hat{H}_0 \rangle_{t_0}/3c^2m\omega_T^2=1.7\cdot 10^{-2}$ (ground state)
and $\omega_{eg}^2\langle \hat{H}_0 \rangle_{t_0}/3c^2m\omega_T^2=0.7$ (thermal state in Doppler limit).
}
\label{fig:excitation+squeezing}

\end{figure}

The influence of squeezing of initially prepared thermal center of mass states on the
excitation probability $P_e(t)$ is illustrated in Fig. \ref{fig:excitation+squeezing} for several scenarios.
These results were derived under the assumption that the condition stated in Eq. (\ref{Doppler_condition}) is satisfied.
If the degree of squeezing becomes too large this condition may be violated and our results no longer apply.
We can take this into account, by using the results from Sec. \ref{SmallLD}.
By using Eq. (\ref{strongconf}), we obtain  
the following probability for exciting the atom for a squeezed thermal center of mass state with $\sin^2(\phi(t_{\text{out}}))=0$ 
\begin{small}
\begin{eqnarray}
P_{e}(t_{\text{out}})&=&1-\frac{\omega_{eg}^{2}}{5c^{2}}\left[\frac{2}{\Gamma^{2}m^{2}}\left(\langle\hat{p}_{z}^{2}(t_{\text{out}})\rangle+2\langle\hat{p}_{x}^{2}(t_{\text{out}})\rangle+2\langle\hat{p}_{y}^{2}(t_{\text{out}})\rangle\right)\right. \nonumber\\
&& -\frac{2}{\Gamma^{2}}\left(\omega_{z}^{2}\langle\hat{z}_{I}^{2}(t_{\text{out}})\rangle+2\omega_{x}^{2}\langle\hat{x}_{I}^{2}(t_{\text{out}})\rangle+2\omega_{y}^{2}\langle\hat{y}_{I}^{2}(t_{\text{out}})\right)\nonumber\\
&&\left.+\langle\hat{z}_{I}^{2}(t_{\text{out}})\rangle+2\langle\hat{x}_{I}^{2}(t_{\text{out}})\rangle+2\langle\hat{y}_{I}^{2}(t_{\text{out}})\rangle\right]+O\left(\frac{\omega_{x,y,z}^3}{\Gamma^3}\right)\nonumber\\
 \end{eqnarray}
\end{small}
for $\left(t_{\text{out}}-t_0\right)\Gamma\gg1$.
For a spherically symmetric trapping potential this simplifies to
\begin{eqnarray}
P_{e}(t_{\text{out}})&=&1-\frac{\omega_{eg}^{2}\langle\psi_{T}|\hat{H}_{0}|\psi_{T}\rangle}{3mc^{2}}\nonumber\\
&&\cdot\left[\left(1-\frac{2\omega_{T}^{2}}{\Gamma^{2}}\right)\frac{1}{\omega_{T}^{2}}e^{-2r\left( t_{\text{out}}\right)}+\frac{2}{\Gamma^{2}}e^{2r \left( t_{\text{out}}\right)}\right]+O\left(\frac{\omega_{T}^3}{\Gamma^{3}}\right)\nonumber\\
&\approx& 1-\frac{\omega_{eg}^{2}\langle\psi_{T}|\hat{H}_{0}|\psi_{T}\rangle}{3mc^{2}}\left[\frac{1}{\omega_{T}^{2}}e^{-2r \left( t_{\text{out}}\right) }+\frac{2}{\Gamma^{2}}e^{2r \left( t_{\text{out}}\right)}\right]\;.\nonumber\\
\end{eqnarray}
By using the above expression, we find that squeezing is increasing the probability for absorbing the photon
as long as
\begin{eqnarray}
r\leq\log\left[\Gamma/(\sqrt{2}\omega_{T})\right]/2
\end{eqnarray}
For the parameters of the experiment described in  \cite{Maiwald2012} ( $\omega_T=2\pi\cdot 480\text{kHz}$, $\Gamma=1.2\cdot 10^8 \text{s}^{-1}$),
we obtain 
$$r\leq 1.7\;.$$
For smaller trapping frequencies or higher decay rates, even larger squeezing parameters $r$ are still beneficial.
Hence, for typical experimental parameters a significant increase of the excitation probability can be achieved.

Although our discussion has concentrated on a spherically symmetric harmonic
trapping potential generalizations to anisotropic cases are straightforward. They
lead to different degrees of squeezing in different directions.

\section{Conclusion \label{Conclusion}}

We have investigated the influence of the center of mass motion of a trapped
two level system
on resonant single photon absorption. In particular, we have concentrated on single photon excitation  by an optimal
photon wave packet which is capable of exciting a two level system at a fixed position almost perfectly. 

It has been demonstrated that the achievable excitation probability depends crucially on the ratio between the time scales
of spontaneous photon emission and absorption on the one hand and of the center of mass motion in the trap on the other hand.
If single photon absorption and emission takes place on a time scale long in comparison with the characteristic time scale of
the center of mass motion in the trap it is the time averaged center of mass motion which determines and limits the achievable
single photon excitation probability. In the opposite limit of fast spontaneous photon emission and absorption it is the
spatial width of the center of mass wave packet at the absorption time which limits the achievable single photon excitation
probability. This latter dependence can be exploited for increasing the achievable excitation probabilities significantly
by squeezing the spatial width of the center of mass wave packet. By modulating the harmonic trapping
frequency appropriately
such a significant squeezing at particular times during the periodic
center of mass motion can be achieved. This way the single photon wave packets considered can achieve highly efficient excitation of a two level
system in free space even if the center of mass motion is only weakly confined and prepared in a thermal state initially.

\begin{acknowledgments}
This work is supported by the DAAD, the BMBF Project Q.com, and by the DFG as part of the CRC 1119 CROSSING.
\end{acknowledgments}

\bibliography{Literatur}
\end{document}